\documentclass[prb,twocolumn,superscriptaddress,showpacs,preprintnumbers,eqsecnum,amsmath,amssymb]{revtex4}
\usepackage{graphicx}
\usepackage{dcolumn}
\usepackage{bm}
\usepackage{color}
\usepackage{epsfig}
\usepackage{psfrag}

\definecolor{darkgreen}{rgb}{0,0.6,0}

\newcommand{\comment}[1]{}
\input xy 
\xyoption{all}

\usepackage{graphicx}
\usepackage{dcolumn}
\usepackage{bm}
\usepackage{color}
\usepackage{epsfig}
\usepackage{psfrag}
\usepackage{amsmath}

\definecolor{darkgreen}{rgb}{0,0.6,0}

\input xy 
\xyoption{all}
\definecolor{darkmagenta}{rgb}{.5,0,.5}
\definecolor{darkgreen}{rgb}{0,0.4,0}
\definecolor{darkred}{rgb}{0.5,0,0}

\begin{document}
\title{
ELECTRON TRANSPORT AND ELECTRON DENSITY
INSIDE QUASI-ONE-DIMENSIONAL DISORDERED CONDUCTORS \\
}


\author{Pier A. Mello} 
\address{Instituto de F\'isica, Universidad  Nacional Aut\'onoma de M\'exico, 04510 Cd. de M\'exico, Mexico}

\author{Miztli Y\'epez}
\address{
Departamento de F\'isica, Universidad Aut\'onoma Metropolitana, Iztapalapa, 
09340 Cd. de M\'exico, Mexico}
\date{\today}

\begin{abstract}
We consider the problem of electron transport across a quasi-one-dimensional disordered multiply-scattering medium, and study the statistical properties of the electron density inside the system. In the physical setup that we contemplate, electrons of a given energy feed the disordered conductor from one end. The physical quantity that is mainly considered is the logarithm of the electron density, $\ln {\cal W}(x)$, since its statistical properties exhibit a self-averaging behavior. We also describe a {\em gedanken} experiment, as a possible setup to measure the electron density. We study analytically and through computer simulations the ballistic, diffusive and localized regimes. We generally find a good agreement between the two approaches. The extension of the techniques that were used in the past to find information outside the sample is done in terms of the scattering properties of the two segments that form the entire conductor on each side of the observation point. The problem is of interest in various other branches of physics, as electrodynamics and elasticity.
\end{abstract}


\maketitle





\



\section{Introduction}
\label{intro}

The problem of the electronic conductance in disordered mesoscopic systems has been of interest for many years (for a review, see Ref. [\onlinecite{mello-kumar}] and references
cited therein).
The equivalence of the electronic conductance expressed in units of the quantum of conductance and the transmittance [\onlinecite{landauer,buettiker,mello-kumar}] has allowed 
studying the electronic transport 
in terms of the scattering properties of the system of interest.
For very low temperatures and small chemical potential difference between the two terminals, the relevant scattering properties are those in the vicinity of the Fermi energy.
That equivalence also allows
many of the predictions of mesoscopic physics and localization theory to apply equally to the transport of quantum and classical waves 
[\onlinecite{cheng1a,cheng1b,cheng1c,cheng1d,cheng1e,cheng1f,cheng1g,cheng1h}]. 

In addition to studies of the conductance and transmission, 
which refer to physical quantities evaluated outside the sample,
the statistics of transport {\em inside} random systems has also been studied for many years [\onlinecite{gazaryan,kohler_papanicolau,
neupane_yamilov,van_tiggelen_et_al_2000-2006,tian_et_al_2010}],
and is of interest in other branches of physics as well, as it is representative of a more general wave-scattering problem in a quasi-one-dimensional (q1D) disordered system:
e.g., an electromagnetic wave traveling in a disordered waveguide
[\onlinecite{cheng-yepez-mello-genack}]
--the interest being in the energy density inside the structure--,
or an elastic wave propagating in a disordered elastic waveguide
[\onlinecite{elastic_waves,elastic_waves_2}].

In this paper, we study the statistical properties of the electron density inside a quasi-one-dimensional multiply-scattering medium.
The system is fed with electrons of a given energy from one end of the disordered conductor and the electron density is evaluated along the conductor and outside.
For 1D systems, this was done recently in 
Refs.  [\onlinecite{cheng-yepez-mello-genack,mello-shi-genack}].
In Ref. [\onlinecite{mello-shi-genack}], the expectation value 
$\langle {\cal{W}}(x)\rangle$ of the intensity ${\cal W}(x)$ at a distance $x$ from the entrance was calculated and compared with computer simulations.
In Ref. [\onlinecite{cheng-yepez-mello-genack}], emphasis was put on the statistics of the logarithm  of the intensity,
$\ln {\cal{W}}(x)$,
which shows interesting scaling properties, in a way similar to the logarithm of the conductance in the conduction problem;
theoretical predictions were compared with computer simulations and microwave experiments.
In the present paper, we address the more complicated problem of the electron density inside q1D
systems supporting more than one propagating mode or open 
channel ($N \ge 1$).

The study of the statistics of electron density in the interior of 
random samples is performed in terms of the scattering properties
of the two segments that form the entire conductor on each side of the observation point.
We apply to each of the two segments, considered to be statistically independent, the maximum-entropy approach (MEA)  
to random-matrix theory [\onlinecite{mello-kumar}] that was used in the past for the entire conductor.
The MEA for the full conductor of 
Ref. [\onlinecite{mello-kumar}]
is a random-matrix theory which leads to a Fokker-Planck equation,
known as the Dorokhov-Mello-Pereyra-Kumar (DMPK) equation
[\onlinecite{dorokhov_82,mpk}],
governing the ``evolution" with sample length $L$ of the 
probability distribution (PD) $p_{L}(M)$ of the system transfer matrix $M$.
In the so-called dense-weak-scattering limit,
the resulting PD for the full system is expected to give results insensitive to microscopic details, depending only on the mean-free-path (MFP) $\ell$.

The multichannel problem is more complicated than the 1D case that was studied in
Refs.  [\onlinecite{cheng-yepez-mello-genack,mello-shi-genack}].
Although we have not been able to solve the problem in its entirety,
we have succeeded in finding a number of partial results, which we consider of sufficient interest to be discussed in the present publication, especially because they may encourage the development of methods towards their experimental verification.
One possible experimental procedure is described in 
Sec. \ref{gedanken experiment}.

Two comments are in order at this point:

i) In a real electrical conduction problem realized by inserting the system between two terminals (reservoirs) at different chemical potentials, the 
{\em electron density} inside the system would have to be calculated by adding the contribution of all incident energies at which electrons are fed by the reservoirs, with a weight given by the Fermi function of the respective reservoir.
As was already mentioned, in the present paper we restrict the analysis to one energy, the more complete calculation being deferred to a later publication.

ii) As was mentioned earlier, the problem studied in this paper may arise in Quantum Mechanics  (QM), describing electronic scattering in a disordered conductor,
or, more generally, in a wave-scattering problem in a q1D disordered waveguide. 
In what follows, we shall refer specifically to the first type of problem and use the QM nomenclature, although the notions to be discussed below are applicable to a more general situation involving wave transport
(see, e.g., Ref. [\onlinecite{beenakker_RMP}], Sec. V-C, for an application to electromagnetic waveguides).

The paper is organized as follows.
In the next section we describe the problem to be developed in the article, 
and give its general mathematical formulation.
With a few exceptions, 
the quantity we shall put emphasis on is the logarithm of the
intensity, $\ln {\cal W}(x)$.
In Sec. \ref{regimes} we discuss the various regimes of interest for this class of systems:
Sec. \ref{ballistic} deals with q1D systems in the ballistic regime
($L/\ell \ll 1$):
we find, for instance, 
a linear dependence of 
$\langle \ln {\cal W}(x) \rangle$ on the position $x$.
In Sec. \ref{N>>1,ball to diff} we study the regime of a large number of open channels, $N \gg 1$ and for 
$0 < s\equiv L/\ell \ll N$, which includes the ballistic and the diffusive regimes:
in this case, 
we were able to find results for the two endpoints of the conductor, $x=0,L$.
Sec. \ref{localized} deals with q1D systems in the localized regime,
$L \gg \xi=(N+1)\ell$, $\xi$ being the localization length.   
For the left endpoint of the conductor, i.e., 
near the incident beam entrance, we find a concrete result for $N=2$ open channels, and a rather general result for the right endpoint.
In Sec. \ref{gedanken experiment} we describe a {\em gedanken} experiment, intended to represent a possible setup to measure the electron density.        
In Sec. \ref{simulations} we verify the theoretical predictions of the previous sections by means of computer simulations.
We give our conclusions in Sec. \ref{concl}.
Various appendices, as well as a number of Supplemental Material (SM) sections,
are included, in order to prove certain specific results
without interrupting the main flow of the paper.

\section{The intensity inside a disordered electronic conductor
with a quasi-1D geometry, supporting $N$ propagating modes}
\label{W(x)N}

\begin{figure}[t]
\centerline{
\includegraphics[width=\columnwidth]
{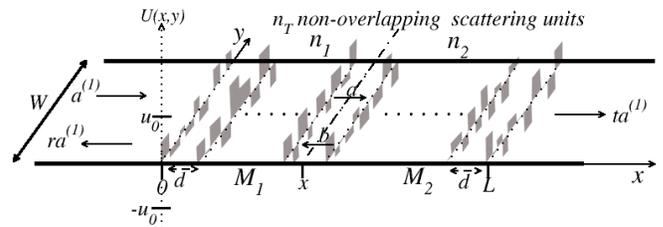}
}
\caption{The scattering problem associated with the q1D disordered conductor described in the text.
The randomness of the potential in the $y$-direction is indicated schematically for one realization of disorder.
For left incidence, the amplitudes of the incident, transmitted and reflected waves at either end of the waveguide are indicated.
We are interested in the intensity a distance $x$ from the entrance.
The first and second segments of the sample contain $n_1$ and $n_2$ 
non-overlapping scattering units, represented schematically 
in the figure; 
the two segments are characterized by the transfer matrices $M_1$, $M_2$, respectively.
The distance between successive random scattering units is $d$.
The maximum value, $|u_0|$, of the random potential for each scattering unit
is indicated.
}
\label{fig1}
\end{figure}

Consider the q1D two-dimensional scattering problem illustrated in Fig. \ref{fig1},
$x$ being the propagation direction and $y$ the transverse direction, respectively;
we contemplate incidence from the left side of the sample at one given energy. 
We shall refer specifically to the electron conduction problem and use the QM nomenclature, although, as was mentioned in the Introduction, our discussion is applicable to a more general problem involving wave transport [\onlinecite{beenakker_RMP}].

The scattering system consists of $n_T$ non-overlapping scattering units;
$n_1$ and $n_2$ of these are located on the left and on the right of the observation abscissa $x$, respectively.
Each scattering unit has a random-potential profile in the $y$-direction, as described more precisely at the beginning of 
Sec. \ref{simulations}.
The whole sample has length $L$, width $W$, and can support $N$ 
{\em open channels}.
Two successive scattering units are a distance $d$ apart (see Fig. \ref{fig1}); 
$d$ will always be taken as $d \ll \lambda$, $\lambda$ being the wavelength.

The scattering problem is defined, as usual, with boundary conditions at $\pm \infty$ in the following way.
We have plane waves incident from the left from $-\infty$ in each of the $N$ open channels, with amplitudes ${a}_n^{(1)}$, $n=1, \cdots, N$.
The effect of the scattering system is to produce reflected waves on the left 
and transmitted waves on the right.

The incident wave function can be written as
\begin{subequations}
\begin{eqnarray}
\psi_{inc}(x, y)
&=&\sum_{n=1}^N a_n^{(1)} \phi_{+}(E_n;x)\chi_n (y)   
\label{psi_inc a}  \\
&=& \sum_{n=1}^N \sqrt{\frac{\mu}{k_n}} \;
|a_{n}^{(1)}|\; e^{i\delta_n} e^{ik_nx} \chi_n (y) \; ,
\label{psi_inc b} 
\end{eqnarray}
where
\begin{eqnarray}
\phi_{\pm}(E_n; x)
&=&\sqrt{\mu} \; \frac{e^{ik_n x}}{\sqrt{k_n}} \;,
\;\;\;   \mu = \frac{m}{2\pi\hbar^2},
\;\;\;  a_n^{(1)} = |a_{n}^{(1)}|\; e^{i\delta_m} \; .       
\nonumber
\\
\label{psi_inc c} \\
E_n &=& \frac{\hbar^2 k_n^2}{2m}, \;\;\;\;
\frac{2mE}{\hbar^2}
= k_n^2 + \left( \frac{n \pi}{W}\right)^2 . 
\end{eqnarray}
\label{psi_inc}
\end{subequations}
Here, $\phi_{\pm}(E_n;x)$ are plane waves with longitudinal momentum
$\hbar k_n$, normalized as delta functions of the energy,
and 
\begin{equation}
\chi_n(y) = \sqrt{\frac{2}{W}}\sin\frac{n\pi y}{W}, 
\;\;\; n=1,\cdots,N \; ,
\label{chi}
\end{equation}
are the transverse wave functions for the $N$ open channels.

The amplitudes of the plane waves incident from the left in each one of the channels form an $N$-dimensional vector, to be indicated as ${a}^{(1)}$.
The amplitudes of the reflected waves will be described by the column vector $r {a}^{(1)}$, $r$ denoting the $N\times N$ reflection matrix, and those of the trasmitted waves on the right will be described by the column vector $t {a}^{(1)}$, $t$ being the $N\times N$ transmission matrix (see Fig. \ref{fig1}).

Inside the conductor we consider a point situated between two successive scattering units, a distance $x$ from the left side of the sample, as illustrated in Fig. \ref{fig1}.
At $x$, the wave function consists of $N$ waves traveling to the right, whose amplitudes form the $N$-dimensional vector $a$, and $N$ waves traveling to the left, with amplitudes forming the vector $b$, i.e.,
\begin{eqnarray}
\psi(x,y)
&=& \sum_{n=1}^N
\big[a_n \phi_{+}(E_n;x) + b_n \phi_{-}(E_n;x) \big] 
\chi_n(y) \; . 
\nonumber
\\
\end{eqnarray}

For one configuration of disorder, we define the {\em linear particle density}, or {\em intensity}, at $x$,
designated as ${\cal W}(x)$, as the particle density 
$|\psi(x,y)|^2$ integrated over the lateral dimension.
Using orthonormality of the transverse wave functions $\chi_n(y)$, we have 
\begin{eqnarray}
{\cal W}(x)
&=& \int_0^{W} |\psi(x,y)|^2 dy 
\nonumber
\\
&=&
\mu \sum_n
\left|a_n \frac{e^{ik_nx}}{\sqrt{k_n}}
+b_n \frac{e^{-ik_nx}}{\sqrt{k_n}}\right|^2 \;  .
\label{W(x)}
\end{eqnarray}

The amplitudes $a_n$, $b_n$ can be calculated from the transfer matrices of
the two portions of the conductor
(a reminder of some general properties of transfer matrices can be found in 
App. \ref{properties of $M$})
and the incident vector of amplitudes 
${a}_n^{(1)}$, as explained in 
App. \ref{calculation of an, bn}
(see also Refs. 
[\onlinecite{cheng-yepez-mello-genack,mello-shi-genack}]).
The intensity appearing in Eq. (\ref{W(x)}) becomes
[see Eq. (\ref{a,b,t 2})]
\begin{eqnarray}
{\cal W}(x)
&=& \mu \sum_n \left|\sum_m
\left[\frac{e^{ik_nx}}{\sqrt{k_n}}(\alpha^{\dagger}_2 t)_{nm} 
-  \frac{e^{-ik_nx}}{\sqrt{k_n}}(\beta^{\dagger}_2 t)_{nm} \right]a_m^{(1)}
\right|^2
\nonumber
\\
\label{W(x) 1 a}
\end{eqnarray}
Here, $\alpha_2, \beta_2$ are the 11 and 12 blocks of the transfer matrix
for the second portion  of the wire, defined in 
App. \ref{calculation of an, bn};
$t$ is the transmission matrix for the full system, defined in
App. \ref{properties of $M$}.
If we assume {\em random phases} $\delta_m$, an average over $\delta_m$ gives [see Eq. (\ref{psi_inc c}), third equation]
\begin{eqnarray}
{\overline{{\cal W}(x)}}^{\delta}
&=& \mu \sum_{n,m} \frac{1}{k_n}\left|
e^{ik_nx}(\alpha^{\dagger}_2 t)_{nm} 
-  e^{-ik_nx}(\beta^{\dagger}_2 t)_{nm}
\right|^2 |a_m^{(1)}|^2 .
\nonumber
\\
\label{W((x) 2}
\end{eqnarray}

At the RHS of the sample, $x=L$, 
we have $\alpha_2=\mathbb{I}_N$ (the $N \times N$ unit matrix),
$\beta_2=0$, 
so that, from Eq. (\ref{W((x) 2}), the intensity is given by
\begin{eqnarray}
{\overline{{\cal W}(L)}}^{\delta}
&=& \mu \sum_{n,m} \frac{1}{k_n}
\left|t_{nm}\right|^2
|a_m^{(1)}|^2 \; .
\label{W(L)}
\end{eqnarray}
Notice that it is not possible to choose the incident probabilities 
$|a_m^{(1)}|^2$ so as to have the intensity on the RHS of 
Eq. (\ref{W(L)}) proportional to the transmittance $T=\sum_{nm}|t_{nm}|^2$,
because $|a_m^{(1)}|^2/k_n$ cannot be made independent of $n$ and $m$.
For these incident probabilities we make the specific choice 
\begin{subequations}
\begin{eqnarray}
|a_m^{(1)}|^2 = \frac{\widetilde{k}}{\mu} \; ,
\label{|am|^2}
\end{eqnarray}
where $\widetilde{k}$ is any quantity with dimensions of a wavenumber.
For later convenience, we choose it as
\begin{eqnarray}
\frac{1}{\widetilde{k}} = \frac{1}{N}\sum_{n=1}^N \frac{1}{k_n} \; ,
\label{k tilde}
\end{eqnarray}
\label{inc prob}
\end{subequations}
i.e., the channel geometric  average of the longitudinal momenta $k_n$.
The {\em intensity} at $x$, Eq. (\ref{W((x) 2}), and at the right end, $x=L$, Eq. (\ref{W(L)}), averaged over the phases $\delta_n$, are then given by
\begin{subequations}
\begin{eqnarray}
{\overline{{\cal W}(x)}}^{\delta}
&=& \sum_{n,m} \frac{\widetilde{k}}{k_n}\left|
e^{ik_nx}(\alpha^{\dagger}_2 t)_{nm} 
-  e^{-ik_nx}(\beta^{\dagger}_2 t)_{nm}
\right|^2   ,
\nonumber
\\
\label{W(x) 2}      \\
{\overline{{\cal W}(L)}}^{\delta}
&=& \sum_{n,m} \frac{\widetilde{k}}{k_n}
\left|t_{nm}\right|^2 .
\label{W(L) 2}
\end{eqnarray}
\end{subequations}
The above choice for $a_m^{(1)}$ gives, for the Quantum Mechanical 
{\em current} through the system, the result
\begin{eqnarray}
\overline{I}^{\delta}
= \frac{\widetilde{k}}{\mu}
\sum_{n,m}|t_{nm}|^2 = \frac{\widetilde{k}}{\mu} T \; ,
\label{g}
\end{eqnarray}
which is proportional to the transmittance $T$.
In contrast, only in an {\em equivalent-channel} (EC) approximation, 
$k_n^{EC} \equiv k_0 \;\; \forall n$, is the intensity at $L$, 
Eq. (\ref{W(L) 2}), equal to the transmittance.
Thus, the choice we have made for the incident probabilities, 
Eqs (\ref{inc prob}),
has the following consequences:
i) the various channels are fed with the same weight, as in 
Ref. [\onlinecite{buettiker}]; ii) the QM current through the system is proportional to the transmittance $T$, 
also as in Ref. [\onlinecite{buettiker}];
iii) in the absence of a system, Eq. (\ref{W(x) 2}) 
(which takes into account that the $k_n$s are $N$ different numbers)
gives 
${\overline{{\cal W}(x)}}^{\delta}/N = 1$,
which is the same result obtained in an EC approximation, 
($k_n^{EC} \equiv \widetilde{k} \;\; \forall n$, 
which implies $\widetilde{k}^{EC}=k_0$),
which is frequently employed in this paper.

For simplicity in writing, in the future we shall omit the bar indicating a phase average, and write ${\cal W}(x)$ for the quantity on the LHS of
Eq. (\ref{W(x) 2}).

\section{The various regimes}
\label{regimes}

\subsection{The ballistic regime 
}
\label{ballistic}

We discuss the intensity ${\cal W}(x)$ defined 
in Eq. (\ref{W(x) 2}), for q1D systems in the ballistic regime, i.e., such that 
\begin{equation}
s\equiv L/\ell \ll 1 \; .
\label{ball reg}
\end{equation}
The intensity ${\cal W}(x)$, evaluated in 
the Supplemental Material (SM) [\onlinecite{SM}], Sec. I,
and given explicitly in Eq. SM(1.4),
simplifies considerably in an EC model, 
$k_n^{EC} \equiv k_0 \;\; \forall n$, giving 
\begin{eqnarray}
\left( \frac{{\cal W}(x)}{N}\right)_{EC}
&=& 
1 - \frac2N  {\rm Re} \left[e^{-2i k_0 x}
{\rm Tr}(v_2^T \sqrt{\lambda_2} \; v_2)\right]
\nonumber
\\
&-&\frac1N {\rm Tr} (\lambda_1)
+\frac1N {\rm Tr} (\lambda_2)
\nonumber
\\
&-&\frac2N {\rm Re} {\rm Tr} [(u_1 \sqrt{\lambda_1} \; u_1^{T})
(v_2^{T} \sqrt{\lambda_2} \; v_2)]
\nonumber
\\
&+&O(\lambda_i^{3/2}) \; .
\label{W(x) ball b}
\end{eqnarray}
We write our expressions employing the polar representation summarized in App. \ref{polar}, in which
the transfer matrix is parametrized in terms of the matrices $\lambda, u, v$:
$\lambda$ is a diagonal $N$-dimensional matrix with non-negative elements 
$\lambda_n$ ($n=1, \cdots N$), and
$u,v$ are arbitrary $N$-dimensional unitary matrices.

The result (\ref{W(x) ball b}) applies to one configuration of disorder.
As shown in SM, Ref. [\onlinecite{SM}], Sec. I, the ensemble average of the logarithm
of the intensity can be written, upon expanding the logarithm, as
\begin{eqnarray}
\left\langle
\ln \frac{{\cal W}(x)}{N}
\right\rangle_{s \ll 1, EC}
&=& \frac{(N-1)(N+2)}{N(N+1)}\frac{L-x}{\ell}
-\frac{x}{\ell}
\nonumber
\\
&+& \left\langle O(\lambda^{3/2}) \right\rangle
+ \cdots \; .
\label{<ln W(x)> ball}
\end{eqnarray}
In the one-channel case, $N=1$, this result reduces to
$-x/\ell$ [independent of the system length (as long as $x<L$)], 
in agreement with the analysis of 
Ref. [\onlinecite{cheng-yepez-mello-genack}].
However, for $N>1$ the result 
(\ref{<ln W(x)> ball}) acquires a dependence on $s=L/\ell$.

This result will be compared with computer simulations in the next section.

\begin{figure}[t]
\centerline{
\includegraphics[width=\columnwidth]
{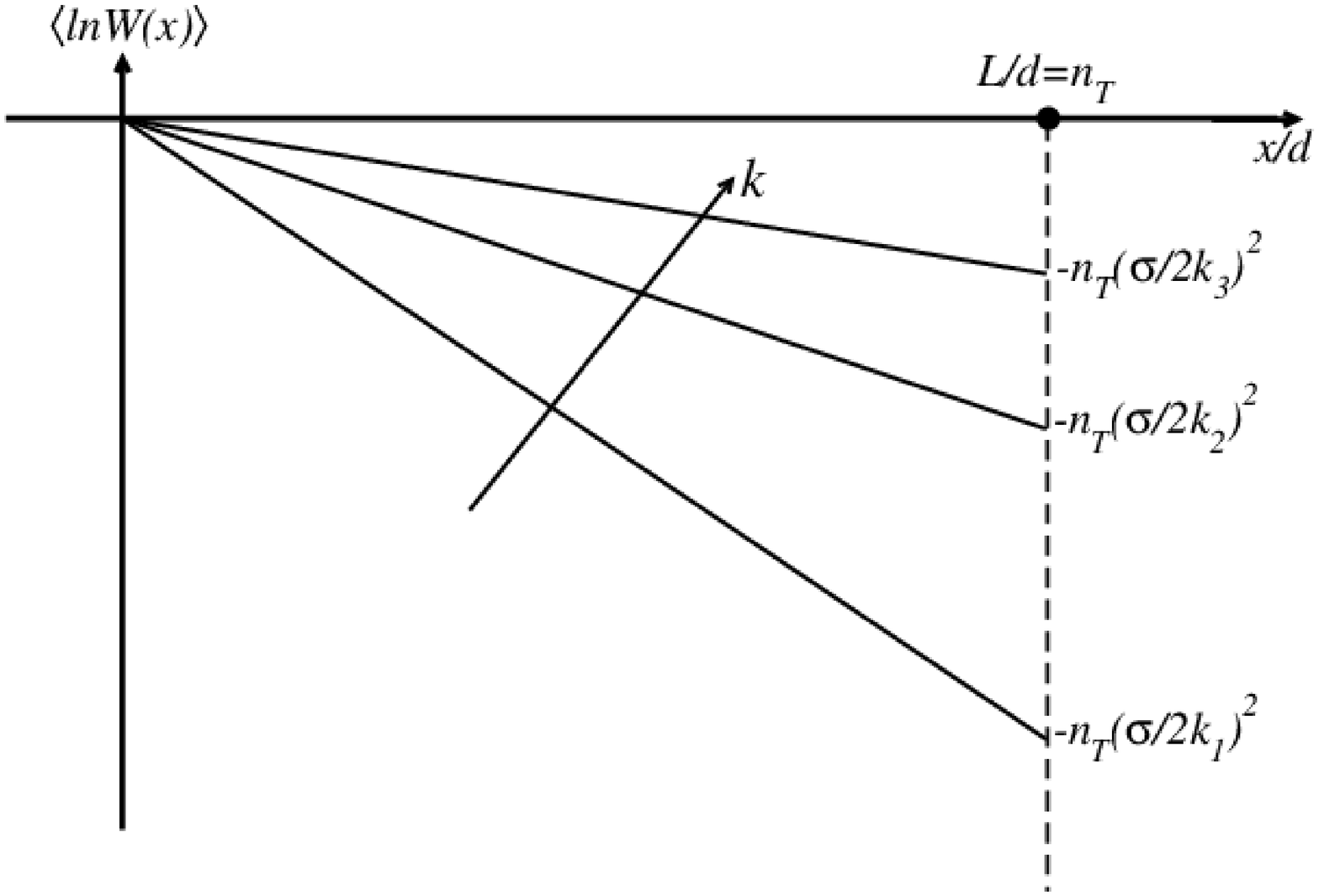}
}
\caption{Behavior of 
$\left\langle
\ln{\cal W}(x)
\right\rangle$, Eq. (\ref{lnW N=1 a}),
as function of $x/d$ for $N=1$, for three values of $k$, 
$k_1 < k_2 <k_3$, showing that 
$\left\langle
\ln{\cal W}(L) 
\right\rangle
=\left\langle
\ln T
\right\rangle \to 0$ as $k$ increases.
}
\label{lnW(x)N1}
\end{figure}

Before concluding this section, we illustrate the dependence on the incident energy of some of the above results.
Consider Eq. (\ref{<ln W(x)> ball}) for $N=1$, which, from Ref. [\onlinecite{cheng-yepez-mello-genack}], is valid $\forall s$; i.e., 
\begin{eqnarray}
\left\langle
\ln{\cal W}(x)
\right\rangle
= -\frac{x}{\ell(k)}.
\label{lnW N=1}
\end{eqnarray}
The $k$ dependence of the MFP can be written as
[\onlinecite{froufe_et_al_pre_2007}]
\begin{subequations}
\begin{eqnarray}
\frac{d}{\ell(k)}
&=&\left(\frac{\sigma}{2k}\right)^2 ,
\end{eqnarray}
where
\begin{eqnarray}
\sigma^2 &=& \langle u_r^2 \rangle, \;\;\; r=1, \cdots, N ;
\label{l(k)}
\end{eqnarray}
\end{subequations}
$d$ is the distance between successive scattering units, and
$u_r$ is the (random) intensity of the $r$-th $\delta$ potential,
with $u_r \in (-u_0, u_0)$ in our model (see Fig. \ref{fig1}).
Then
\begin{subequations}
\begin{eqnarray}
\left\langle
\ln{\cal W}(x)
\right\rangle
= -\frac{x}{d} \left(\frac{\sigma}{2k}\right)^2 \; .
\label{lnW N=1 a}
\end{eqnarray}
In particular,
\begin{eqnarray}
\left\langle
\ln{\cal W}(L) 
\right\rangle
=\left\langle
\ln T
\right\rangle
= - n_T \left(\frac{\sigma}{2k}\right)^2 \; .
\label{lnW N=1 b}
\end{eqnarray}
\label{lnW N=1}
\end{subequations}
This last result tends to 0 as $k$ increases
(since $T \to 1$), as it should;
it is illustrated in Fig. \ref{lnW(x)N1}.

\subsection{The case of a large number of open channels.
From the ballistic to the diffusive regime}
\label{N>>1,ball to diff}

We now consider the case when the number of open channels is very large, 
\begin{subequations}
\begin{eqnarray}
&& 
 N \gg 1 ,
\label{N>>1} 
\end{eqnarray}
and for  
\begin{eqnarray}
&&  
0 < s=L/\ell \ll N ,
\label{s<<N}
\end{eqnarray}
\label{N>>1 and s<<N}
\end{subequations}
conditions that cover the ballistic and diffusive regimes; 
we restrict the analysis to the two end points of the sample, i.e., $x=0,L$, for which we have explicit results.

We start from Eq. (\ref{W(0) 1}), which gives the intensity at the {\em left end of the sample} and compute its expectation value. 
In the DMPK approach, $ \langle r_{nm} \rangle = 0$,
and $\langle \left|r_{nm}\right|^2\rangle$
is independent of $n,m$, so that
\begin{subequations}
\begin{eqnarray}
\langle r_{nm} \rangle &=& 0 \; , 
\label{<r>=0}   \\
\langle |r_{nm}|^2 \rangle
&=& 
\frac{\langle R \rangle}{N^2}
= \frac{N - \langle T \rangle}{N^2}
= \frac1N - \frac{\langle T \rangle}{N^2} \; ,
\label{<|r|2>}
\end{eqnarray}
\end{subequations}
in terms of the reflectance
$R=\sum_{n,m}|r_{nm}|^2$
and the transmittance
$T=\sum_{n,m}|t_{nm}|^2$.
Since in the present regime the ensemble expectation value of the transmittance can be approximated as 
[\onlinecite{mello-kumar}]
\begin{equation}
\langle T \rangle \approx \frac{N}{1+s}  \; ,
\label{T for N gg 1 1}
\end{equation}
Eq. (\ref{W(0) 1}) gives, on average,
\begin{subequations}
\begin{eqnarray}
\left\langle\frac{{\cal W}(0)}{N}\right\rangle
&=& 1 + \frac1N \left(\sum_{nm} \frac{\widetilde{k}}{k_n}\right) 
\left(\frac1N \frac{s}{1+s}\right) \; ,
\label{<W0> N gg 1 a} \\
&=& \frac{1+2s}{1+s} 
\approx \left\{
\begin{array}{ccc}
1+s, &{\rm for}& s \ll 1 \; ,  \\
2-\frac1s, &{\rm for}& 1 \ll s \ll N \; .
\end{array}
\right.\; ,
\nonumber
\\
\label{<W0> N gg 1 b} 
\end{eqnarray}
\label{<W0> N gg 1}
\end{subequations}
so for the logarithm of this expression we obtain
\begin{equation}
\ln \left\langle\frac{{\cal W}(0)}{N}\right\rangle
= \ln \frac{1+2s}{1+s} 
\approx 
\left\{
\begin{array}{ccc}
s, &{\rm for}& s \ll 1 \; , \\
\ln 2 -\frac{1}{2s}, &{\rm for}& 1 \ll s \ll N \; .
\end{array}
\right.
\label{ln<W0> N gg 1 theo}
\end{equation}

At the {\em right end of the sample}, $x=L$, Eq. (\ref{W(L) 2}) gives
\begin{equation}
\frac{{\cal W}(L)}{N}
= \frac1N\sum_{n,m} \frac{\widetilde{k}}{k_n}
\left|t_{nm}\right|^2 .
\label{W(L) 3}
\end{equation}
In the DMPK approach we have
\begin{subequations}
\begin{eqnarray}
\langle |t_{nm}|^2 \rangle 
&=& \frac{\langle T \rangle}{N^2} \; , 
\label{Tnm a}  \\
&\approx& \frac1N \frac{1}{1+s} \; ,
\label{Tnm b}
\end{eqnarray}
\end{subequations}
where we used the result (\ref{T for N gg 1 1}), valid 
in the regime defined by Eqs. (\ref{N>>1 and s<<N}).
Substituting in Eq. (\ref{W(L) 3}), we then find
\begin{subequations}
\begin{eqnarray}
\left\langle\frac{{\cal W}(L)}{N}\right\rangle
&=& \frac{1}{1+s}
\label{<WL> N gg 1}
\\
&=& \left\{
\begin{array}{lcc}
1-s+\cdots , &{\rm for}& s\ll 1 \; , \\
\frac1s - \frac{1}{s^2} + \cdots, &{\rm for}& 1 \ll s \ll N  \; ,
\end{array}
\right.
\nonumber
\\
\ln \left\langle\frac{{\cal W}(L)}{N}\right\rangle
&=& - \ln (1+s)
\label{ln <WL> N gg 1}
\\
&=& \left\{
\begin{array}{ccc}
-s + \cdots, &{\rm for}& s\ll 1 \; , \\
-\ln s + \cdots , &{\rm for}& 1 \ll s \ll N \; .
\end{array}
\right.
\nonumber
\end{eqnarray}
\end{subequations}

These results will be compared with computer simulations in the next section.

\subsection{The localized regime}
\label{localized}

In this regime one finds $N$ characteristic lengths 
 $\xi_n = \frac{(N+1)\ell}{n}$,
$n=1, \cdots, N$.
The conductance is dominated by 
$\xi_1 = (N+1)\ell \equiv \xi$, which we call the localization length $\xi$ [\onlinecite{beenakker_RMP}].
Here we consider the length of the system to fulfill
\begin{equation}
L \gg \xi=(N+1)\ell \; .
\label{local reg}
\end{equation}
We restrict our analysis to the left and right ends of the sample, i.e., $x=0, L$. 

In the localized regime, from Eq. (\ref{W(0) 2})
we can approximate, in lowest order 
($\lambda_n \gg 1, \;\;n = 1,\cdots N$), the logarithm of the intensity 
at the {\em left end} as
\begin{eqnarray}
\left( \frac{{\cal W}(0)}{N} \right)_{s \gg N}
&=&  1 - \frac{1}{N}\sum_{n =1}^{N}\frac{\widetilde{k}}{k_n}
\sum_{a =1}^{N}(v_{an}v_{an} + v_{an}^{*}v_{an}^{*}) 
\nonumber 
\\
&+& \frac{1}{N}\sum_{n,m=1}^{N}
\frac{\widetilde{k}}{k_n}
\left(\sum_{a=1}^N  v_{an}v_{am}\right) 
\left(\sum_{b=1}^N  v_{bn}^{*}v_{bm}^{*}\right) \; .
\nonumber
\\
\label{W(0) loc}
\end{eqnarray}
We notice that in the present approximation,
$\lambda_n \gg 1$, any function of ${\cal W}(0)/N$ 
has no $\lambda$ dependence left.
The statistics of such a function will thus be independent of 
the system length $s=L/\ell$.
In particular, the tendency of 
$\left\langle \ln [{\cal W}(0)/N] \right\rangle$ to become independent of $L/\ell$ as this parameter increases is verified in the computer simulation for two open channels, $N=2$, to be discussed later in relation with 
Fig. \ref{<lnW0> N2 1}. 
This result is similar to that found in the 1D case for the average
of $\ln {\cal W}(x)$, which, for fixed $x$,
gives $-x/\ell$ and is independent of $L/\ell$ for arbitrary values of this parameter 
(of course, by definition, $0<x<L$; see Ref. [\onlinecite{cheng-yepez-mello-genack}], Fig. 1 and related discussion).

As an illustration of the approximation we have used, the reflection coefficient
$R_{mn} = |r_{mn}|^2$ of Eq. (\ref{Rnm polar}) and the total reflection coefficient $R_n=\sum_m R_{mn}$ would be given, in the same approximation, by
\begin{subequations}
\begin{eqnarray}
R_{mn} 
&\approx & \sum_{a,b=1}^N v_{an} v_{am} (v_{bn} v_{bm})^* 
\\
R_n &=& \sum_{m=1}^N R_{mn} 
\approx \sum_{a,b=1}^N  v_{an} v_{bn}^*
\sum_{m=1}^N  v_{am} v_{bm}^*
\nonumber
\\
&=&\sum_{a,b=1}^N  v_{an} v_{bn}^* \delta_{ab}
= \sum_{a=1}^N v_{an} v_{an}^* 
=1 \;  ,
\end{eqnarray}
\end{subequations}
meaning that a wave incident in channel $n$ is fully reflected into all $N$ backward channels, with no transmission left.
This is the crudest approximation to localization.

If we adopt an EC approximation,
$k_n^{EC} \equiv k_0 \;\; \forall n$,
Eq. (\ref{W(0) loc}) reduces to
\begin{eqnarray}
\left( \frac{{\cal W}(0)}{N} \right)_{s \gg N, EC}
&=& 2 - \frac{1}{N}\sum_{a,n=1}^{N}
(v_{an}v_{an} + v_{an}^{*}v_{an}^{*}) \; ,
\nonumber
\\
\label{W(0) loc EC}
\end{eqnarray}
The expectation value of the logarithm of (\ref{W(0) loc EC}) is then
\begin{eqnarray}
&&\left\langle \ln \frac{{\cal W}(0)}{N} \right\rangle_{s \gg N, EC}
= 
\label{lnW(0) loc EC}
\\
&&\left\langle 
\ln \left[1 - \frac{1}{2N}\sum_{a,n=1}^{N}
(v_{an}v_{an} + v_{an}^{*}v_{an}^{*})\right]
\right\rangle_0
+ \ln 2 \; ,
\nonumber
\end{eqnarray}
where the index $0$ indicates an average over the matrices $v$, distributed according to the invariant measure of the group $U(N)$
[\onlinecite{mello-kumar}].
This is evaluated in SM, Ref. [\onlinecite{SM}], Sec. 2, for $N=2$.
From Eq. (2.6b)SM, we have 
\begin{equation}
\left\langle \ln \frac{{\cal W}(0)}{2} \right\rangle_{s \gg 2, EC}
= - 0.122351 +\ln 2  = 0.570796.
\label{lnW(0)> N2 1}
\end{equation}

At the {\em right end} of the sample, $x=L$, the intensity is given by Eq. (\ref{W(L) 2}).
Only in an EC approximation it coincides with the transmittance, i.e.,
\begin{subequations}
\begin{eqnarray}
{\cal W}(L)
= \sum_{m,n=1}^N |t_{mn}|^2  
= T ,
\label{W(L) EC a}
\end{eqnarray}
which, in the polar representation, is
\begin{eqnarray}
\frac{{\cal W}(L)}{N}
= \frac1N \sum_{a=1}^N \frac{1}{1+\lambda_a}.
\label{W(L) EC polar}
\end{eqnarray}
\label{W(L) EC}
\end{subequations}
In the localized regime, Eqs. (\ref{W(L) EC}) give
\begin{subequations}
\begin{eqnarray}
\left\langle
\ln \frac{{\cal W}(L)}{N}
\right\rangle_{s \gg N, EC}
&=& \langle \ln T \rangle_{s \gg N} - \ln N  \\
& \approx & -2 \frac{L}{(N+1)\ell} - \ln N \; 
\nonumber
\\.
\label{<lnWL/N> EC}
\end{eqnarray}
\end{subequations}

These results will be compared with computer simulations in the next section.

\subsection{Measuring the electron density}
\label{gedanken experiment}

Before closing this section, we describe a {\em gedanken} experiment, intended to represent a possible setup to measure the electron density that we have been studying in the previous sections.
For simplicity, we shall deal with 1D conductors.

\begin{figure}[t]
\centerline{
\includegraphics[width=\columnwidth]
{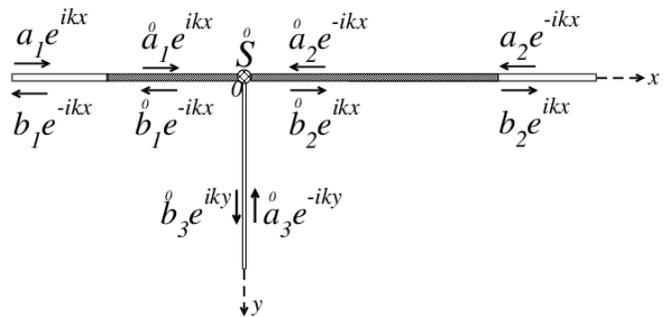}
}
\caption{
The setup used to measure the electron density at the point $x=0$ of 
our 1D system.
The symbol at $x=0$ represents the junction used to join the ``probe" wire to the main wire, described by the $\stackrel{o}{S}$ matrix of Eqs. (\ref{S0}), 
}
\label{probe setup 1}
\end{figure}

\begin{figure}[t]
\centerline{
\includegraphics[width=\columnwidth]
{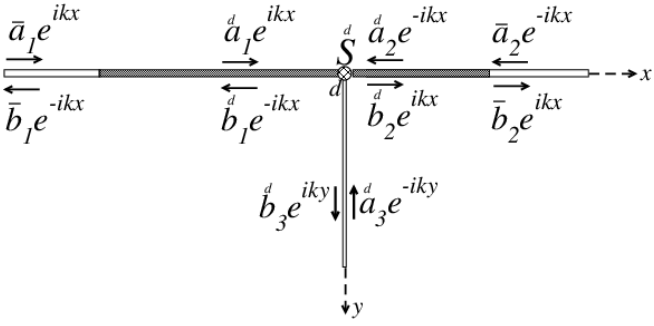}
}
\caption{
The same setup as in Fig. \ref{probe setup 1}, designed to measure the electron density at the point $x=d$ of our 1D system.
}
\label{probe setup 2}
\end{figure}

The main idea of the procedure is to introduce an additional wire, as a ``probe" to measure the electron density at some point along the main wire,
as illustrated in Figs. \ref{probe setup 1} and \ref{probe setup 2}.
This setup, devised by B\"uttiker for a number of other purposes
[\onlinecite{buettiker}], is similar to the one used in actual electromagnetic experiments, where an antenna is introduced at the point of interest to measure the electromagnetic energy density [\onlinecite{cheng-yepez-mello-genack}].
A ``junction"
joins the probe to the main wire.

Consider first the probe being introduced at the point defined as $x=0$
in Fig. \ref{probe setup 1}.
The incoming waves along the two portions of the main wire
at $x=0$ and along the probe are designated as
$\stackrel{o}{a_1}{\rm e}^{ikx}$, $\stackrel{o}{a_2}{\rm e}^{-ikx}$
and $\stackrel{o}{a_3}{\rm e}^{-iky}$,
and the corresponding outgoing waves as   
$\stackrel{o}{b_1}{\rm e}^{-ikx}$, $\stackrel{o}{b_2}{\rm e}^{ikx}$
and $\stackrel{o}{b_3}{\rm e}^{iky}$.
They are related by the junction scattering matrix $\stackrel{o}{S}$, for which we take the simple model introduced in Ref. [\onlinecite{buettiker}], as
\begin{subequations}
\begin{eqnarray}
\stackrel{o}{S}
\left[
\begin{array}{cc}
\stackrel{o}{a_1} \\
\stackrel{o}{a_2} \\
\stackrel{o}{a_3}
\end{array}
\right]
&=&
\left[
\begin{array}{cc}
\stackrel{o}{b_1} \\
\stackrel{o}{b_2} \\
\stackrel{o}{b_3}
\end{array}
\right] \; ,
\label{S0 1}  \\
\stackrel{o}{S}
&=&
\left[
\begin{array}{ccc}
\alpha & \beta & \sqrt{\epsilon}   \\
\beta  & \alpha  &  \sqrt{\epsilon}  \\
\sqrt{\epsilon}  & \sqrt{\epsilon}  & -(\alpha + \beta)
\end{array}
\right]\; ,
\label{S0 2}
\end{eqnarray}
\label{S0}
\end{subequations}
with $\alpha$, $\beta$ real and given by
\begin{subequations}
\begin{eqnarray}
\alpha 
= -\frac12 \left[ 1-\sqrt{1-2 \epsilon}  \right] 
\label{alpha}  \\
\beta
= \frac12 \left[ 1+\sqrt{1-2 \epsilon}  \right].
\label{beta}
\end{eqnarray}
\end{subequations}
The coupling strength is described by the real parameter $\epsilon \in[0, 1/2]$.
Actually, we shall take $\stackrel{o}{a_3}=0$, i.e., no incoming waves in the probe lead, and $\stackrel{o}{b_3}= \stackrel{o}{t_3}$, i.e., the transmitted amplitude along the probe lead.

Eqs. (\ref{S0}) then give
\begin{equation}
\stackrel{o}{a_1} + \stackrel{o}{a_2}
= \frac{ \stackrel{o}{t_3} }{\sqrt{\epsilon}} \; .
\label{eqn for t3 1}
\end{equation}
In the limit as the coupling strength $\epsilon \to 0$,
we have $\stackrel{o}{a_1} \to \stackrel{o}{a}$,
$\stackrel{o}{a_2} \to \stackrel{o}{b}$, 
in the notation used in the previous sections for the amplitudes of the wave traveling to right and left at the point of interest (see Fig. \ref{fig1}), 
so that 
\begin{equation}
\left[ \stackrel{o}{a}{\rm e}^{ikx} + \stackrel{o}{b}{\rm e}^{-ikx}\right]_{x=0}
= \lim_{\epsilon \to 0} 
\frac{  \stackrel{o}{t_3} } {\sqrt{\epsilon}}.
\label{eqn for t3 1}
\end{equation}
The square of the expression on the LHS is the intensity at the point $x=0$,
and is thus expressed in terms of the transmission coefficient 
$\stackrel{o}{T_3}=|\stackrel{o}{t_3}|^2$ {\em measured} along the probe lead, and the coupling strength $\epsilon$, the latter being a characteristic of the junction.

Repeating this experiment over a large sample, we can find the required statistical properties of the intensity.

We now consider the probe being introduced at $x=d$, as in 
Fig. \ref{probe setup 2}.
The incoming waves along the two portions of the main wire at $x=d$ and along the probe are designated as
$\stackrel{d}{a_1}{\rm e}^{ikx}$, $\stackrel{d}{a_2}{\rm e}^{-ikx}$
and $\stackrel{d}{a_3}{\rm e}^{-iky}$,
and the corresponding outgoing waves as   
$\stackrel{d}{b_1}{\rm e}^{-ikx}$, $\stackrel{d}{b_2}{\rm e}^{ikx}$
and $\stackrel{d}{b_3}{\rm e}^{iky}$.
They are related by the junction scattering matrix $\stackrel{d}{S}$,
which can be expressed in terms of the previous junction scattering matrix $\stackrel{o}{S}$ through the translation matrix $D(kd)$ as
(see Ref. [\onlinecite{mello-kumar}])
\begin{subequations}
\begin{eqnarray}
\stackrel{d}{S}
&=& D(kd) \stackrel{o}{S} D(kd) \; ,
\label{Sd vs S0 a} \\
D(kd)
&=& \left[
\begin{array}{ccc}
{\rm e}^{ikd} & 0 & 0   \\
0  & {\rm e}^{-ikd}  &  0  \\
0  & 0  & 1
\end{array}
\right]\; .
\label{S0 2}
\label{transl matrix D}
\end{eqnarray}
\label{Sd vs S0}
\end{subequations}

In the limit as the coupling strength $\epsilon \to 0$,
we have $\stackrel{d}{a_1} \to \stackrel{d}{a}$,
$\stackrel{d}{a_2} \to \stackrel{d}{b}$, 
in the notation used in the previous sections for the amplitudes of the wave traveling to right and left at the point of interest (see Fig. \ref{fig1}), 
so that 
\begin{equation}
\left[ \stackrel{d}{a}{\rm e}^{ikx} + \stackrel{d}{b}{\rm e}^{-ikx}\right]_{x=d}
= \lim_{\epsilon \to 0} 
\frac{  \stackrel{d}{t_3} } {\sqrt{\epsilon}}.
\label{eqn for t3 d 1}
\end{equation}
The square of the expression on the LHS is the intensity at the point $x=d$,
and is expressed in terms of the transmission coefficient 
$\stackrel{d}{T_3}=|\stackrel{d}{t_3}|^2$ {\em measured} along the probe lead and the coupling strength $\epsilon$, the latter being a characteristic of the junction.

Again, repeating this experiment over a large sample, we can find the required statistical properties of the intensity.

\section{Computer simulations}
\label{simulations}

To check the theoretical results of the previous sections, we present a number of computer simulations of random q1D systems supporting $N$ propagating modes, in which the disordered potential is a random function of position.
We use the model employed in Ref. [\onlinecite{froufe_et_al_pre_2007}],
in which the scattering units consist of thin potential slices, idealized as equidistant delta potentials 
($d$ being their separation), perpendicular to the longitudinal direction of the conductor, the variation of the potential in the transverse direction being random. 
This is illustrated in Fig. \ref{fig1}.
The sets of parameters defining a given slice are taken to be statistically independent from those of any other slice and identically distributed. In the dense-weak-scattering limit, in which the potential strength of the slices is very weak and their linear density is very large, so that the resulting mean free paths are fixed, the corresponding statistical properties of the full system depend only on the mean free paths and on no other property of the slice distribution. 
\begin{figure}[t] 
\centering 
\includegraphics[width=\columnwidth]
{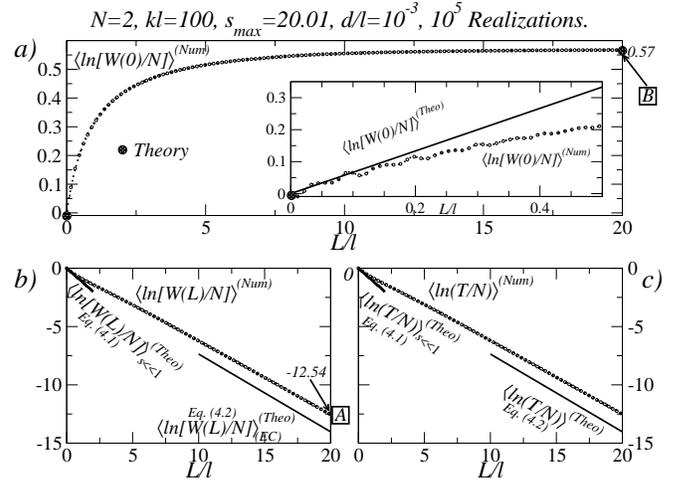} 
\caption{
\footnotesize{
Computer simulations for $N=2$ open channels and some theoretical results for the quantities shown in the various panels.
The values of the parameters used in the simulations are given in the top part of the figure.
{\em a)} Numerical results for $\langle \ln ({\cal W}(0)/N) \rangle$, as a function of $s=L/{\ell}$.
The inset shows the ballistic regime in more detail.
The simulation point marked \fbox{$B$} in the figure has to be compared with point \fbox{$B$}
in Fig. \ref{<lnWx> N2 ball}b for $s=L/\ell=20.01$.
{\em b)} Numerical results for 
$\langle \ln ({\cal W}(L)/N) \rangle$ 
and theoretical results for $s \ll 1$ and $s \gg 1$,
as functions of $L/{\ell}$.
The simulation point marked \fbox{$A$} in the figure has to be compared with point \fbox{$A$} in Fig. \ref{<lnWx> N2 ball}b.
{\em c)} Numerical results for $\langle \ln (T/N) \rangle$
and theoretical results for $s \ll 1$ and $s \gg 1$,
as functions of $L/\ell$.
}}
\label{<lnW0> N2 1}
\end{figure}

We first consider the {\em left end} of the sample, $x=0$.
The quantity $\langle \ln ({\cal W}(0) /N) \rangle$ was calculated numerically by a computer simulation for $N=2$ open channels, {\em from the ballistic to the localized regime}, as function of $s=L/\ell$, as can be seen in Fig. \ref{<lnW0> N2 1}{\em a}.
In the {\em ballistic} regime, corresponding to $s \ll 1$,
the result is in excellent agreement with 
$(2/3)L/\ell$,
as predicted by Eq. (\ref{<ln W(x)> ball}), 
which we reproduce here
\begin{eqnarray} 
\left\langle
\ln \frac{{\cal W}(x)}{N}
\right\rangle_{s \ll 1}
&\approx&  \frac{(N-1)(N+2)}{N(N+1)}\frac{L-x}{\ell}
-\frac{x}{\ell}
+ \cdots \; ,
\nonumber
\\
\label{W(x) ball 3}
\end{eqnarray}
evaluated at $x=0$.
The result is not $L$ independent, as it is for $N=1$.
In the {\em localized} regime, $s \gg N=2$, both the simulation 
(point marked \fbox{B} in the figure) and theory tend asymptotically to 
$\approx 0.57$, a constant independent of $s$.

\begin{figure}[t]
\centering
\includegraphics[width=\columnwidth]
{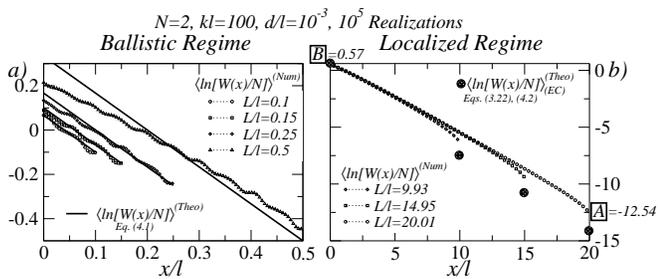}
\caption{
\footnotesize{
Computer simulations of the profile $\langle \ln {\cal W}(x) /N \rangle$ 
for $N=2$ open channels and some theoretical predictions.
{\em a)} Numerical results for the ballistic regime and slightly beyond, and comparison with theory, Eq. (\ref{W(x) ball 3}),
shown as full lines.
{\em b)} Numerical results for the localized regime.
Theory is given only for $x=0$ and $x=L$,  and is indicated by big dots.
The origin of the discrepancy is probably the same as discussed in the text in relation with Fig. \ref{<lnW0> N2 1}. 
}
}
\label{<lnWx> N2 ball}
\end{figure}

Numerical results
for the {\em right end} of the sample
are given in panel \ref{<lnW0> N2 1}{\em b}
for $\langle \ln ({\cal W}(L)/N) \rangle$, and 
in panel \ref{<lnW0> N2 1}{\em c} for $\langle \ln (T/N) \rangle$.
In panel \ref{<lnW0> N2 1}{\em b} we also show, for $s \ll 1$, 
the 
outcome of theoretical analysis
[Eq. (\ref{W(x) ball 3})]
for $\left\langle \ln ({\cal W}(L)/N) \right\rangle_{s \ll 1}$, 
which is seen to agree well with the 
simulation, and in panel \ref{<lnW0> N2 1}{\em c} 
for $\left\langle \ln (T/N) \right\rangle_{s \ll 1}$.
Panels \ref{<lnW0> N2 1}{\em b} and \ref{<lnW0> N2 1}{\em c} also show the 
theoretical prediction of 
Eq. (\ref{<lnWL/N> EC}) for $s \gg N=2$, which we reproduce here
\begin{eqnarray}
\left\langle
\ln \frac{{\cal W}(L)}{N}
\right\rangle_{s \gg N, EC}
&=& \left\langle \ln \frac{T}{N} \right\rangle_{s \gg N}
\nonumber
\\
&\approx& -2 \frac{L}{(N+1)\ell} - \ln N \; .
\label{<lnWL/N> EC 2}
\end{eqnarray}
There is a difference between the 
simulations in 
panels \ref{<lnW0> N2 1}{\em b} and \ref{<lnW0> N2 1}{\em c}, not noticeable in the figure, due to the two channels not having the same longitudinal momentum $k_n$:
compare ${\cal W}(x)$ in Eq. (\ref{W(L) 2}) with $T$ in Eq. (\ref{g}).
We see a discrepancy between 
theory and simulations 
for $s \gg N=2$.
A useful example to understand this discrepancy 
is to compare the behavior of $\langle \ln R \rangle$ 
with that of $\langle \ln T \rangle$ as $s$ increases.
Although one always has $R+T=N$, the quantities $\langle \ln R \rangle$ and $\langle \ln T \rangle$ behave very differently.
As $s$ increases, $\langle \ln R \rangle$ tends to $\ln N$ for both theory and simulations.
On the other hand, $\langle \ln T \rangle$ is not bounded as $s$ increases: it tends to $-\infty$, 
so the difference between theory and simulations cannot be assessed a priory.
In the present case there is a difference in value and slope
of $\langle \ln T \rangle$ vs $s$
between theory and simulations, as seen in Fig. \ref{<lnW0> N2 1}{\em c}:
the two curves get farther apart for larger values of $s$.
In order to be more specific, one would need an improved theory beyond DMPK, such as the theories 
developed in 
Refs. [\onlinecite{chalker,muttalib,suslov}].

Fig. \ref{<lnWx> N2 ball} shows, in panel {\em a}, the profile 
$\langle \ln ({\cal W}(x)/N) \rangle$ for $N=2$ open channels as function of $x$ for various values of $s$, for the {\em ballistic} 
regime, $s\ll 1$, and slightly beyond, $s\sim 1$.

The theoretical prediction of Eq. (\ref{<ln W(x)> ball}), 
reproduced above 
in Eq. (\ref{W(x) ball 3}),
agrees well with the computer simulations
up to $s\sim 0.3$.
The simulations show oscillations as function of $x/\ell$, which are not reproduced by the DMPK treatment.
On the other hand, these interference oscillations are described well
by an analysis based on Born's approximation, up to second order
[\onlinecite{yepez-saenz}].
Fig. \ref{<lnWx> N2 ball}{\em b} shows the behavior of the profile 
$\langle \ln ({\cal W}(x) /N) \rangle$ for $N=2$ channels as a function of $x$, for various values of $s$ in the {\em localized} regime.
For $x=0$, the theoretical result (i.e., 0.57) 
is given in Eq. (\ref{lnW(0)> N2 1});
point \fbox{B} ($=0.57$, too) is the result of simulation.
For $x=L$, theory is that of Eq. (\ref{<lnWL/N> EC 2}),
while \fbox{A} is the simulation result.
In the localized regime, as $x$ goes from the left to the right edge of the sample, $x=0$ to $x=L$,
the discrepancy between theory and simulations increases:
it is imperceptible for $x=0$ and very noticeable for $x=L$,
as seen in Fig. \ref{<lnWx> N2 ball}{\em b}.
See the above discussion in relation with Figs. \ref{<lnW0> N2 1}{\em b,c}.

\begin{figure}[t]
\centerline{
\includegraphics[width=\columnwidth]
{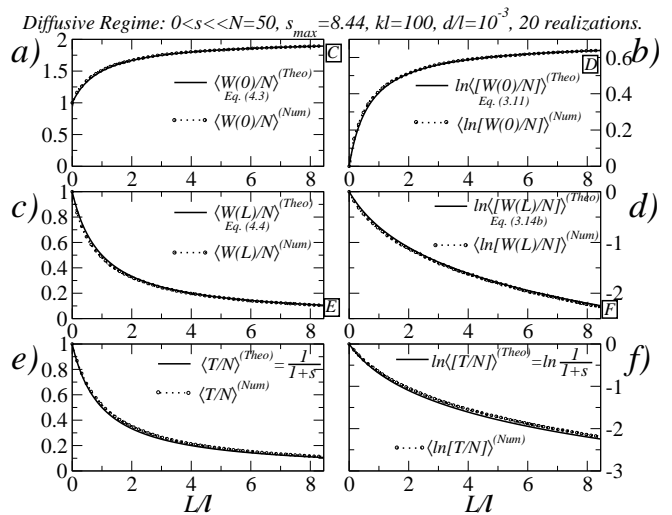}
}
\caption{
\footnotesize{
Computer simulations (dotted lines)
and theoretical results (continuous lines) for the quantities shown in the various panels.
The equations in Sec. \ref{N>>1,ball to diff} from which the latter are taken are indicated in the text.
Results correspond to $0 < s < N$, and go from the {\em ballistic} to approximately the {\em diffusive} regime.
We discuss in the text the quality of the agreement between the two calculations.
Points marked as \fbox{$C$}, \fbox{$D$}, \fbox{$E$}, and \fbox{$F$} 
represent numerical results which
correspond to the ones marked similarly in 
Fig. \ref{ball and diffusive Oct18 a}.
}
}
\label{ball and diffusive Oct18}
\end{figure}

We compare, in Fig. \ref{ball and diffusive Oct18}, theoretical results with computer simulations for the regime in which the number of open channels is very large, $N \gg 1$, and for 
$0 < s=L/\ell \ll N$.
The theoretical description is given in Sec. \ref{N>>1,ball to diff}.
In the left panels, the theoretical results are taken from 
Eq. (\ref{<W0> N gg 1 b}) (which we reproduce here)
for panel {\em a}, i.e.,
\begin{eqnarray}
\left\langle\frac{{\cal W}(0)}{N}\right\rangle
&=& \frac{1+2s}{1+s} 
\approx 
\left\{
\begin{array}{ccc}
1+s, &{\rm for}& s \ll 1 \; ,  \\
2-\frac1s, &{\rm for}& 1 \ll s \ll N \; ,
\end{array}
\right.
\nonumber
\\
\label{<W0> N gg 1 bb} 
\end{eqnarray}
from Eq. (\ref{<WL> N gg 1}) (which we also reproduce here) for panel 
{\em c}, i.e.,
\begin{eqnarray}
\left\langle\frac{{\cal W}(L)}{N}\right\rangle
&=& \frac{1}{1+s} 
=\left\{
\begin{array}{lcc}
1-s+\cdots , &{\rm for}& s\ll 1 \; , \\
\frac1s - \frac{1}{s^2} + \cdots, &{\rm for}& 1 \ll s \ll N  \; ,
\end{array}
\right. 
\nonumber
\\
\label{<WL> N gg 1 1}            
\end{eqnarray}
and from $1/(1+s)$ for panel {\em e}.
In the panels on the right, the simulations represent the average of a logarithm, while
the theoretical results are the logarithm of an average:
the latter are taken from Eq. (\ref{ln<W0> N gg 1 theo}) for panel 
{\em b}, 
Eq. (\ref{ln <WL> N gg 1}) for panel {\em d},
and $-\ln (1+s)$ for panel {\em f}.
That the two calculations give almost the same results is an indication
of the self-averaging property for the logarithm of the various quantities considered (see Ref. [\onlinecite{cheng-yepez-mello-genack}], Supplemental Material).

\begin{figure}[t]
\centerline{
\includegraphics[width=\columnwidth]
{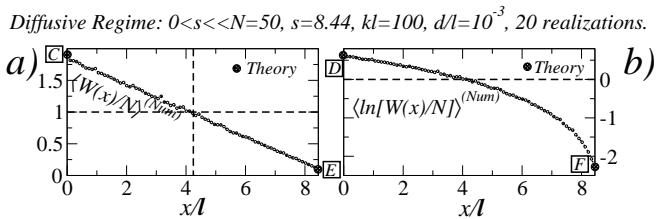}
}
\caption{
\footnotesize{
For the same data as in Fig. \ref{ball and diffusive Oct18} 
and for $s= 8.44$, which is the maximum value of $s$ in 
Fig. \ref{ball and diffusive Oct18} (so we are approximately in the diffusive regime), computer simulations for the profile of
{\em a)}  $\langle {\cal W}(x)/N\rangle$ as a function of $x/\ell$;
{\em b)} $\langle \ln ({\cal W}(x)/N)\rangle$ as a function of $x/\ell$.
Theoretical results for $\langle {\cal W}(x)/N\rangle$
and $\ln\langle {\cal W}(x)/N\rangle$ for $x=0$ and $x=L$ are indicated as big dots.
Points marked as \fbox{C}, \fbox{D}, \fbox{E}, \fbox{F} 
represent numerical results which
correspond to the ones marked similarly in 
Fig. \ref{ball and diffusive Oct18}.
}
}
\label{ball and diffusive Oct18 a}
\end{figure}

Fig. \ref{ball and diffusive Oct18 a} shows computer simulations for the profiles
$\langle {\cal W}(x)/N\rangle$ 
and $\langle \ln ({\cal W}(x)/N)\rangle$
as function of $x/\ell$, for the same data as 
Fig. \ref{ball and diffusive Oct18},
$N=50$ and for $s=8.44$ (the maximum value of $s$ in Fig. 
\ref{ball and diffusive Oct18}), so we are approximately in the {\em diffusive} regime.
The equivalent of panel {\em a} for the 1D case
[\onlinecite{mello-shi-genack}]
would be an s-shaped curve going from 2 to zero, being antisymmetrical with respect to the intersection of the two dashed lines shown in panel {\em a}.
The equivalent of panel {\em b} for the 1D case
[\onlinecite{cheng-yepez-mello-genack}] would be a 45-degree straight line starting from zero on the left.

\section{Summary and Conclusions}
\label{concl}

In this paper we studied the statistical properties of the electron density inside a q1D multiply-scattering disordered electric conductor
which may support more than one
propagating mode ($N \ge 1$).

The physical quantity that was mainly considered was
the logarithm of the electron density,
$\ln {\cal W}(x)$, 
since its statistical properties possess a self-averaging behavior, as was explained, for $N=1$, in the supplemental material to
Ref. [\onlinecite{cheng-yepez-mello-genack}].
The theoretical analysis in the present work was based on the DMPK model
[\onlinecite{mello-kumar}], and the results were compared with computer simulations.

We first addressed the case in which the system is in the ballistic regime.
We encountered, in Eq. (\ref{<ln W(x)> ball}), a {\em linear} dependence of
$\langle \ln {\cal W}(x) \rangle$ on  the position $x$.
Also, for $x=0$, we found that the result is {\em linear} in $s=L/\ell$.
For one open channel, $N=1$, 
$\langle \ln {\cal W}(x) \rangle = -x/\ell$ ($x \in (0,L))$ is independent of the 
system length $L$, a remarkable property shown in Ref. [\onlinecite{cheng-yepez-mello-genack}];
on the other hand, a dependence on $L$ appears in 
$\langle \ln {\cal W}(x) \rangle$
if $N>1$. 
Theoretical results are in good agreement with simulations.
However, we should point out that the latter exhibit oscillations as function of $x/\ell$, which are missed by the DMPK treatment. 
In contrast, these oscillations are reproduced by an analysis based on Born's approximation, as studied in Ref. [\onlinecite{yepez-saenz}].
 
We then dealt with systems possessing a large number of propagating modes, 
or open channels, $N \gg 1$, when the parameters $s$ and $N$ fulfill
$0 < s\equiv L/\ell \ll N$, this inequality including the ballistic and the diffusive regimes.
For $\langle {\cal W}(x) / N\rangle$ at $x=0$ 
we obtained the result 
(\ref{<W0> N gg 1 bb}), that goes from 1 to 2  as $s$ goes from 0 to 
the regime $1 \ll s \ll N$.
For $\langle {\cal W}(x) / N\rangle$ at $x=L$
we found the typical $1/(1+s)$ behavior of the average transmittance, Eq. (\ref{<WL> N gg 1 1}).
The theoretical results for this regime are in very good accordance with the computer simulations.
We encountered evidence of a self-averaging property, judging from the good agreement between the expectation value of the logarithm of the electron density and the logarithm of the expectation value, the latter being the quantity that was amenable to a theoretical calculation.

We finally analyzed the q1D systems in the localized regime.
In an equivalent-channel approximation, we obtained a rather general expression for $\langle \ln {\cal W}(x) \rangle$ for $x=0$,
i.e., near the incident beam entrance on the left side of the sample,
Eq. (\ref{lnW(0) loc EC}),
which we could compute explicitly for two open channels, $N=2$, 
Eq. (\ref{lnW(0)> N2 1}),
and compare with computer simulations.
We also derived a rather general expression for $\langle \ln {\cal W}(x) \rangle$ for $x=L$, as in this case of ECs it is directly connected with the transmittance.
We noticed that, in the localized regime, as $x$ goes from the left to the right edge of the sample, $x=0$ to $x=L$,
the discrepancy between theory and simulations increases,
becoming quite noticeable for $x=L$, as exhibited in Fig. \ref{<lnWx> N2 ball}{\em b}.
We conjectured that such a discrepancy is due to the approximations involved in the DMPK equation,
like the isotropy assumption, whose influence is more apparent in the localized than in the diffusive regime; such an
assumption has been relaxed in a number of models, like those presented in
Refs. [\onlinecite{chalker,muttalib,suslov}].

The multichannel problem is more complicated than the 1D case that was studied in Refs.
[\onlinecite{cheng-yepez-mello-genack,mello-shi-genack}].
As a matter of fact, because of technical difficulties we have not been able to compute theoretically, for any number of propagating modes $N > 1$ 
(as was done
in Ref. [\onlinecite{cheng-yepez-mello-genack}]
for $N=1$): 
i) the profile $\langle \ln {\cal W}(x) \rangle$ for arbitrary positions $x$ and  for arbitrary $s=L/\ell$, except for the particular positions and for the regimes $s$ mentioned above;
on the other hand, we reported on computer simulations for arbitrary values of these parameters;
ii) the profile of the variance of $ \ln {\cal W}(x)$, again for any $x$ and  
$s$.
These problems remain open for future studies.

As was indicated in the Introduction, a real electrical conduction problem is modeled by placing the system between two reservoirs at different chemical potentials
[\onlinecite{landauer,buettiker}].
The electrons fed by the reservoirs at all energies, weighted by the respective Fermi function, would then contribute to the electron density inside the system.
In contrast, in the present paper we restricted the analysis to 
the system being fed with electrons of a given energy from one end of the disordered conductor, and the electron density was then evaluated for that energy along the conductor and outside.
The more complete calculation will be performed in a future publication.

We also described a {\em gedanken} experiment for 1D electric conductors, intended to represent a possible experimental setup to measure the electron density that was studied in this paper.
This possible suggested experiment would complement those that have been performed
in the classical domain, as in electromagnetic disordered waveguides 
[\onlinecite{cheng-yepez-mello-genack}] and in elastic media 
[\onlinecite{elastic_waves,elastic_waves_2}].

\appendix


\section{Technical details}
\label{technical details}

\subsection{Some properties of the transfer matrix}
\label{properties of $M$}

By definition, the transfer matrix $\boldsymbol{M}$ for the full system relates the amplitudes on the 
RHS, $a^{(2)}, b^{(2)}$,
to those on the LHS, $a^{(1)}, b^{(1)}$, as
\begin{eqnarray}
\boldsymbol{M} \left[
\begin{array}{c}
{a}^{(1)}    \\
{b}^{(1)} 
\end{array}
\right]
= \left[
\begin{array}{c}
{a}^{(2)}    \\
{b}^{(2)} 
\end{array}
\right]     \; ;
\end{eqnarray}
$\boldsymbol{M}$ is a $2N\times 2N$ matrix, with the structure
\begin{eqnarray}
\boldsymbol{M} = 
\left[
\begin{array}{cc}
\alpha   & \beta \\
\beta^{*} & \alpha^{*}
\end{array}
\right] \; ,
\label{M_i}
\end{eqnarray}
satisfying the properties of time-reversal and flux conservation, as
\begin{subequations}
\begin{eqnarray}
\boldsymbol{M}^{\dagger} \boldsymbol{\Sigma}_z \boldsymbol{M} 
= \boldsymbol{\Sigma}_z \; , \\
\boldsymbol{M}^* = \boldsymbol{\Sigma}_x \boldsymbol{M}
\boldsymbol{\Sigma}_x   \; ,
\end{eqnarray}
\end{subequations}
respectively.
Here, $\boldsymbol{\Sigma}_z$ and $\boldsymbol{\Sigma}_x$ are the $2N \times 2N$ generalization of Pauli matrices ${\bf \sigma}_z$ and ${\bf \sigma}_x$.
The $N$-dimensional blocks of the $\boldsymbol{M}$ matrix are related to the reflection and transmission matrices $r$, $t$ for left incidence and 
$r'$, $t'$ for right incidence as 
\begin{subequations}
\begin{eqnarray}
&& r = -(\alpha^*)^{-1} \beta^* , \hspace{1cm} t' =  (\alpha^*)^{-1} ,   \\
&& t= (\alpha^{\dagger})^{-1}  , \hspace{1.5cm}  r' =  \beta (\alpha^*)^{-1} \; .
\end{eqnarray}
\end{subequations}

\subsection{Calculation of the amplitudes $a_n$, $b_n$ appearing in Eq. (\ref{W(x)})}
\label{calculation of an, bn}

The transfer matrices of the two portions of the wire are denoted by
\begin{equation}
\boldsymbol{M}_i = 
\left[
\begin{array}{cc}
\alpha_i   & \beta_i \\
\beta_i^{*} & \alpha_i^{*}
\end{array}
\right] , \;\;\;\;\; i=1,2.
\label{M_i}
\end{equation}
From the definiton of the transfer matrix, we have
\begin{equation}
\boldsymbol{M}_2
\left[
\begin{array}{c}
a    \\
b 
\end{array}
\right]
=
\left[
\begin{array}{c}
t  \; a^{(1)}  \\
0 
\end{array}
\right] \; ,
\label{a,b,t 1}  
\end{equation}
so that the amplitudes between the two portions are given by
\begin{equation}
\left[
\begin{array}{c}
{a}    \\
{b}  
\end{array}
\right]
=
\left[
\begin{array}{r}
(\alpha_2^{\dagger} t) \; {a}^{(1)} \\
-(\beta_2^{\dagger} t) \; {a}^{(1)} 
\end{array}
\right] ,
\hspace{5mm}
t=\frac{1}{\alpha^{\dagger}}
= \frac{1}{\alpha_1^{\dagger}\alpha_2^{\dagger}  + \beta_1^{T}\beta_2^{\dagger}} \; .
\label{a,b,t 2}  
\end{equation}
This result is used in Eq. (\ref{W(x) 1 a}).



\subsection{The polar representation}
\label{polar}

From Ref. [\onlinecite{mello-kumar}], a transfer matrix 
$\boldsymbol{M}$ can be represented as
\begin{equation}
\boldsymbol{M} =
\left[
\begin{array}{cc}
u  &  0 \\
0  &   u^{*}
\end{array}
\right]
\left[
\begin{array}{cc}
\sqrt{1+\lambda}   & \sqrt{\lambda}  \\
\sqrt{\lambda}   &  \sqrt{1+\lambda} 
\end{array}
\right]
\left[
\begin{array}{cc}
v  &  0 \\
0  &   v^{*}
\end{array}
\right] \; .
\label{polar 1}
\end{equation}
Here, $u$ and $v$ are $N \times N$ arbitrary unitary matrices;
$\lambda$ is an $N \times N$ diagonal matrix with elements 
$\lambda_n \ge 0$, $n=1, \cdots N$.

The reflection matrix $r$ can be expressed as
\begin{equation}
r = -v^{T}\sqrt{\frac{\lambda}{1+ \lambda}} v
\label{r}
\end{equation}
These results are used in Sec. \ref{regimes}.

\section{Exact result for $x=0$}
\label{exact x=0}

For $x=0$, using Eq. (\ref{a,b,t 2}) 
(see also Ref. [\onlinecite{mello-kumar}] for the relation between the transfer matrix and the $S$ matrix)  we have
\begin{subequations}
\begin{eqnarray}
\alpha_2 
&=&\alpha = \frac{1}{t^{\dagger}} \; ,
\hspace{20mm}
\alpha_2^{\dagger} t = \mathbb{I}_N \; , 
\hspace{15mm} a = a^{(1)},
\nonumber
\\
\\
\beta_2 
&=& \beta = -\frac{1}{t^{\dagger}} r^{*} \; ,
\hspace{1cm} 
-\beta_2^{\dagger} t = r^T = r \; , 
\hspace{8mm} b = r a^{(1)} \; .
\nonumber
\\
\end{eqnarray}
\label{a,b x=0}
\end{subequations}
Eq. (\ref{W(x) 2}) then gives
\begin{equation}
\frac{{\cal W}(0)}{N}
= 1 + 
\frac{1}{N}\sum_{n=1}^{N} \frac{\widetilde{k}}{k_n} 
(r_{nn}+ r_{nn}^{*})
+\frac{1}{N}\sum_{n,m=1}^{N}
\frac{\widetilde{k}}{k_n} |r_{nm}|^2 \; .
\label{W(0) 1}
\end{equation}
Using the polar representation parameters introduced in 
App. \ref{polar}, we can write the reflection matrix $r$, its matrix elements, and the reflection coefficients as
\begin{subequations}
\begin{eqnarray}
r &=& -v^T \sqrt{\frac{\lambda}{1+\lambda}} v \; , 
\label{r polar}   \\
r_{nm}
&=& - \sum_{a=1}^{N} v_{an}v_{am} \sqrt{\frac{\lambda_a}{1+\lambda_a}} \; , 
\label{rnm polar}  \\
R_{nm}& \equiv &|r_{nm}|^2 
= \sum_{a,b=1}^N 
\sqrt{\frac{\lambda_a}{1+\lambda_a}}
\sqrt{\frac{\lambda_b}{1+\lambda_b}} \;
v_{an} v_{am} v_{bn}^* v_{bm}^*.
\nonumber
\\
\label{Rnm polar}
\end{eqnarray}
\end{subequations}
Substituting in Eq. (\ref{W(0) 1}), ${\cal W}(0)/N$ becomes
\begin{eqnarray}
&&\frac{{\cal W}(0)}{N}
=
1 - \frac{1}{N}\sum_{n=1}^N \sum_{a=1}^{N}\frac{\widetilde{k}}{k_n}
\sqrt{\frac{\lambda_a}{1+\lambda_a}}
(v_{an}v_{an} + v_{an}^{*}v_{an}^{*}) 
\nonumber  \\
&& +\frac{1}{N}\sum_{n,m=1}^{N}
\frac{\widetilde{k}}{k_n}
\sum_{a,b=1}^N\sqrt{\frac{\lambda_a}{1+\lambda_a}}
\sqrt{\frac{\lambda_b}{1+\lambda_b}}
(v_{an}v_{am} v_{bn}^{*}v_{bm}^{*}) \; .
\nonumber
\\
\label{W(0) 2}
\end{eqnarray}
This result is valid {\em for any regime} $L/\ell$.
It is used in Secs. \ref{N>>1,ball to diff} and \ref{localized}.



\end{document}


\title{SUPPLEMENTAL MATERIAL \\
for the paper: \\
ELECTRON TRANSPORT AND ELECTRON DENSITY
INSIDE DISORDERED QUASI-ONE-DIMENSIONAL CONDUCTORS \\
}


\author{Pier A. Mello} 
\address{Instituto de F\'isica, Universidad  Nacional Aut\'onoma de M\'exico, 04510 Cd de M\'exico, Mexico}

\author{Miztli Y\'epez}
\address{
Departamento de F\'isica, Universidad Aut\'onoma Metropolitana, Iztapalapa, 
09340 Cd. de M\'exico, Mexico}
\date{\today}


\maketitle



\section{Calculation of the intensity in the ballistic regime}
\label{ball. regime}

The submatrix $\alpha$ of the transfer matrix (Eq. (A2) of the main text) for the full system is given in terms of the polar parameters of the two individual sections of the wire as
\begin{equation}
\alpha
= \left(u_2 \sqrt{1+\lambda_2}v_2\right)
\left(u_1 \sqrt{1+\lambda_1}v_1\right)
+ \left(u_2 \sqrt{\lambda_2}v_2^*\right) 
\left( u_1^*\sqrt{\lambda_1}v_1\right)  \; .
\label{alpha(1,2)}
\end{equation}
The transmission matrix $t$, written in Eq. (A4b) of the main text as
$t=1/\alpha^{\dagger}$, is given,
in the ballistic regime, up to terms $\lambda_i^{3/2}$, by
\begin{equation}
t= u_2v_2 \left[
I_N 
-\frac12 u_1 \lambda_1 u_1^{\dagger}  
-\frac12 v_2^{\dagger} \lambda_2 v_2  
- (u_1\sqrt{\lambda_1}u_1^T)(v_2^T\sqrt{\lambda_2}v_2)
+O(\lambda_i^2)
\right]
u_1v_1 \; .
\label{t ball reg}
\end{equation}
The terms $\alpha_2^{\dagger}t$ and $\beta_2^{\dagger}t$, 
appearing in Eq. (2.9a) of the main text, then take the form
\begin{subequations}
\begin{eqnarray}
\alpha_2^{\dagger}t
&=& u_1v_1 -\frac12(u_1\lambda_1 v_1)
-(u_1 \sqrt{\lambda_1} u_1^T)(v_2^T \sqrt{\lambda_2} v_2)u_1 v_1 +O(\lambda_i^2)
\label{alpha2+t} \\
\beta_2^{\dagger}t
&=& (v_2^T \sqrt{\lambda_2} v_2)u_1 v_1
+O(\lambda_i^{3/2}) \; .
\label{beta2+t}
\end{eqnarray}
\end{subequations}

Substituting expressions (\ref{alpha2+t}) and (\ref{beta2+t}) in 
Eq. (2.9a) of the text, 
we find
\begin{eqnarray}
{\cal W}(x)
= \sum_n 
\frac{\widetilde{k}}{k_n}
\Bigg\{
1 - 2 {\rm Re}\left[e^{-2ik_nx}(v_2^T \sqrt{\lambda_2} v_2)\right]_{nn}
-(u_1 \lambda_1 u_1^{\dagger})_{nn}
+(v_2^{\dagger} \lambda_2 v_2)_{nn} 
\nonumber   \\
-2{\rm Re} \left[(u_1 \sqrt{\lambda_1} u_1^{T})
(v_2^{T} \sqrt{\lambda_2} v_2) \right]_{nn} 
+O(\lambda_i^{3/2})
\Bigg\}
\label{W(x) ball b}
\end{eqnarray}

In an equivalent-channel (EC) approximation, 
$k_n^{EC} \equiv k_0 \;\; \forall n$,
an ensemble average of the logarithm of the intensity can be written, upon expanding the logarithm, as
\begin{subequations}
\begin{eqnarray}
\left\langle \ln \frac{{\cal W}(x)}{N}\right\rangle_{s \ll 1, EC}
= 
&&  -\frac2N {\rm Re}\left[e^{-2i k_0 x}
{\rm Tr}\left\langle v_2^T \sqrt{\lambda_2} \; v_2\right\rangle_{s_2}\right]
\label{a} \\
&&  -\frac1N {\rm Tr} \left\langle \lambda_1\right\rangle_{s_1}  
\label{b} \\
&&  +\frac1N {\rm Tr} \left\langle \lambda_2 \right\rangle_{s_2}
\label{c}   \\
&& -\frac2N {\rm Re} {\rm Tr} \left[
\left\langle u_1 \sqrt{\lambda_1} \; u_1^{T} \right\rangle_{s_1}
\left\langle v_2^{T} \sqrt{\lambda_2} \; v_2 \right\rangle_{s_2} \right]
\label{d}   \\
&& -\frac{1}{2N^2}
\left\langle
\left[
e^{-2i k_0 x}
{\rm Tr} \left(v_2^T \sqrt{\lambda_2} \; v_2 \right) 
+
e^{2i k_0 x}
{\rm Tr} \left(v_2^{\dagger} \sqrt{\lambda_2} \; v_2^{*}\right)
\right]^2
\right\rangle_{s_1,s_2}
\label{e} \\
&& +O(\lambda_i^{3/2}) \; .
\nonumber
\end{eqnarray}
\end{subequations}
In Eq. (\ref{a}) we have
\begin{subequations}
\begin{eqnarray}
&& {\rm Tr}\left\langle 
v_2^T \sqrt{\lambda_2} \; v_2
\right\rangle_{s_2} 
=\sum_{ab} \left\langle \sqrt{(\lambda_2)_{b}}\right\rangle_{s_2} \;
\left\langle (v_2)_{ba}\; (v_2)_{ba} \right\rangle_0
=0  \; ,
\end{eqnarray}
the index $0$ indicating an average on the group $U(N)$, evaluated with the invariant measure.
In Eq. (\ref{b}) we have
\begin{eqnarray}
&&  -\frac1N {\rm Tr} \left\langle \lambda_1\right\rangle_{s_1}  
 = -\frac1N N \frac{L_1}{\ell}
 = -\frac{L_1}{\ell} 
 = -\frac{x}{\ell}
\end{eqnarray}
In Eq. (\ref{c}) we have 
\begin{eqnarray} 
&&  \frac1N {\rm Tr} \left\langle \lambda_2\right\rangle_{s_2}  
 = \frac1N N \frac{L_2}{\ell}
 = \frac{L_2}{\ell}
= \frac{L-x}{\ell}
\end{eqnarray}
In Eq. (\ref{d}), by the same argument as in (\ref{a}), we find 
\begin{eqnarray}  
&& \left\langle u_1 \sqrt{\lambda_1} \; u_1^{T} \right\rangle_{s_1}
\left\langle v_2^{T} \sqrt{\lambda_2} \; v_2 \right\rangle_{s_2}
=0 
\end{eqnarray}
In Eq. (\ref{e}) we have, upon expanding the square, the following terms
\begin{eqnarray}
&& \left\langle \left[ {\rm Tr} 
\left( v_2^T \sqrt{\lambda_2} \; v_2 \right)\right]^2\right\rangle 
= 0    \\
&& \left\langle \left[ {\rm Tr} 
\left( v_2^{\dagger} \sqrt{\lambda_2} \; v_2^{*} \right)\right]^2\right\rangle 
= 0   \\
&&  \left\langle {\rm Tr} \left(v_2^T \sqrt{\lambda_2} \; v_2 \right)
{\rm Tr} \left(v_2^{\dagger} \sqrt{\lambda_2} \; v_2^* \right)
\right\rangle
= \frac{2}{N+1}N\frac{L_2}{\ell} \; ,
\end{eqnarray}
which give, for (\ref{e})
\begin{eqnarray}
(\ref{e}) 
=-\frac{2}{N(N+1)}\frac{L_2}{\ell} \; .
\end{eqnarray} 
\end{subequations}
Putting the various terms together, we find the result (3.3) of the text.

\section{The localized regime}
\label{app loc regime}

In order to compute the expectation value appearing 
in Eq. (3.21) of the main text, we first define the unitary, symmetric matrix $s=vv^T$ 
and its eigenvalues ${\rm e}^{i \theta_a}$.
We can write
\begin{subequations}
\begin{eqnarray}
\sum_{a,n=1}^{N} v_{an}v_{an} 
&=&{\rm  Tr} (vv^T)  \\
&\equiv& {\rm Tr} s 
= \sum_{a=1}^N {\rm e}^{i \theta_a}  \\
{\rm with:} \hspace{1cm} ss^{\dagger} &=& I, \;\;\;\;\;  s=s^T \; .
\end{eqnarray}
\end{subequations}
If $v$ is distributed according to the 
{\em invariant measure of the group} $U(N)$, $s$ is distributed according to the {\em invariant measure of unitary, symmetric $N$-dimensional matrices}, and its eigenvalues according to the distribution [\onlinecite{dyson}, \onlinecite{mehta}]
\begin{equation}
p(\theta_1, \cdots, \theta_N)
= C_N \prod_{1=a<b}^{N}
\left| {\rm e}^{i \theta_a} - {\rm e}^{i \theta_b} \right| \; .
\label{P(thetas) N 1}
\end{equation}
The expectation value in Eq. (3.21) of the main text
can thus be written as
\begin{subequations}
\begin{eqnarray}
\left\langle \ln \frac{{\cal W}(0)}{N} \right\rangle_{s \gg N, EC}
&=& \left\langle \ln \left(1 - \frac{1}{N}\sum_{a=1}^{N}
\cos \theta_a \right) \right\rangle_0 
+ \ln 2 \\
&=& \int_0^{2\pi}  \cdots \int_0^{2\pi} \ln 
\left(1 - \frac{1}{N}\sum_{a=1}^{N}
\cos \theta_a \right) 
p(\theta_1, \cdots, \theta_N) d\theta_1 \cdots d\theta_N 
+ \ln 2 \; .
\label{<lnW(0)> N 1}
\nonumber \\
\end{eqnarray}
\end{subequations}
Here, the index $0$ indicates that the average over $v$ is computed according to the invariant measure of the unitary group $U(N)$. 

As an example, Eq. (\ref{<lnW(0)> N 1}) gives, for $N=1$,
\begin{equation}
\left\langle \ln {\cal W}(0) \right\rangle_{s \gg 1} 
=\int_0^{2\pi} 
\ln (1-\cos \theta) \frac{d \theta}{2\pi} 
+ \ln 2
= - \ln 2 + \ln 2 = 0 \; ,
\label{<lnW(0)> N1 loc 1}
\end{equation}
verifying the result of Ref. [\onlinecite{cheng-yepez-mello-genack}].

For $N=2$, Eq. (2.2) gives 
\begin{equation}
p(\theta_1,\theta_2)
= \frac{1}{8 \pi \sqrt{2}}
\sqrt{1-\cos(\theta_1-\theta_2)} \; .
\label{P(theta1,2) 1}
\end{equation}
The integral needed in Eq. (\ref{<lnW(0)> N 1}) was evaluated numerically, with the result
\begin{subequations}
\begin{eqnarray}
&&\left\langle \ln \left(1-\frac{\cos \theta_1 + \cos \theta_2}{2}\right) \right\rangle_0 
\nonumber \\
&& \hspace{2cm}= \int_0^{2\pi} d\theta_1  \int_0^{2\pi } d\theta_2
\ln \left(1-\frac{\cos \theta_1 + \cos \theta_2}{2}\right)
p(\theta_1,\theta_2) \\
&& \hspace{2cm}= -0.122351 \; ,
\label{<ln(1-(c+c)/2)>} 
\end{eqnarray}
\end{subequations}
as quoted in Eq. (3.22) of the main text.

